\documentclass[twocolumn,prb,showpacs,superscriptaddress,amsmath,citeautoscript]{revtex4}
\usepackage{graphicx}
\usepackage{bm}
\usepackage{amssymb}

\begin{document}

\title{Onset of metallic behavior in strained (LaNiO$_3$)$_n$/(SrMnO$_3$)$_2$ superlattices}

\author{S.\ J.\ May} 
\affiliation{
Materials Science Division, Argonne National Laboratory, Argonne, IL 60439\\}
\author{T.\ S.\ Santos}
\affiliation{
Center for Nanoscale Materials, Argonne National Laboratory, Argonne, IL 60439\\}
\author{A.\ Bhattacharya}
\email{anand@anl.gov}
\affiliation{
Materials Science Division, Argonne National Laboratory, Argonne, IL 60439\\}
\affiliation{
Center for Nanoscale Materials, Argonne National Laboratory, Argonne, IL 60439\\}

\date{\today}


\begin{abstract}
(LaNiO$_3$)$_n$/(SrMnO$_3$)$_2$ superlattices were grown using ozone-assisted molecular beam epitaxy.  \textit{In situ} reflection high energy electron diffraction and x-ray scattering has been used to characterize the structural properties of the superlattices, which are strained to the SrTiO$_3$ substrates.  The superlattices exhibit excellent crystallinity and interfacial roughness of less than one unit cell.  A metal-insulator transition is observed as $n$ is decreased from 4 to 1.  Analysis of the transport data suggests an evolution from gapped insulator ($n$=1) to hopping conductor ($n$=2) to metal ($n$=4) with increasing LaNiO$_3$ concentration.
\end{abstract}

\maketitle
\section{Introduction}
 Transition metal oxides exhibit an array of collectively ordered states, including magnetism, ferroelectricity and superconductivity, making these materials promising candidates for future applications.\cite{Salamon,Schlom,Maekawa04}  While the manganites, titanates and cuprates have been investigated extensively, nickelate heterostructures have received less attention.\cite{Nikolaev,Padhan03,Lee06,Granada}  Bulk LaNiO$_3$ (LNO) does not exhibit ordering phenomena such as (anti)ferromagnetism or superconductivity.  Instead, LNO is a paramagnetic metal in which the Ni $3d^7$ electrons ($t_{2g}^{6}e_{g}^{1}$) hybridize with the O $2p$ states to form the conduction band.  However, the properties of LNO, like many complex oxides, can be altered significantly through the formation of heterostructures with different oxide materials.  For instance, the double perovskite La$_2$MnNiO$_6$, equivalent to a LaMnO$_3$/LaNiO$_3$ superlattice with a (111) growth direction, is a ferromagnetic insulator ($T_C$= 280 K).  The Ni in La$_2$MnNiO$_6$ exhibits a $3d^8$ configuration, leading to a ferromagnetic superexchange interaction between the Ni$^{2+}$ and Mn$^{4+}$.\cite{Sanchez02,Rogado05}  Presumably, the same mechanism is responsible for the ferromagnetism reported in a LaMnO$_3$/LaNiO$_3$ superlattice ($T_C$= 210 K) deposited on a SrTiO$_3$ (001) substrate.\cite{Tanaka00}    
 
 Recent theoretical work suggests that strain and electronic confinement effects may give rise to magnetic order in LaNiO$_3$ layers where the Ni retains its bulk $3d^7$ electronic configuration.  Dobin \emph{et al.} \cite{Dobin03} predicted the emergence of ferromagnetism in LNO films under tensile strains ($c<a$) greater than 3.5 $\%$.  Chaloupka and Khaliullin suggested antiferromagnetism and high-$T_C$ superconductivity may be stabilized in LNO-based superlattices, assuming the LNO layers are under tensile strain and charge transfer along the $c$-axis between LNO layers is suppressed by insulating layers of other perovskite oxides.\cite{Chaloupka08}  While confinement of carriers to the LNO layers is critical to this prediction, there have been few experimental reports on transport and its dimensionality in short period LNO-based superlattices such as those proposed by Chaloupka and Khaliullin.  Padhan and Budhani reported activated insulating behavior in (LaNiO$_3$)$_n$/(La$_{0.7}$Ca$_{0.3}$MnO$_3$)$_{10}$ superlattices when $n <$ 5, a unexpected result as bulk LNO and La$_{0.7}$Ca$_{0.3}$MnO$_3$ are metallic.\cite{Padhan03}  They attributed this insulating behavior to structural and/or magnetic disorder at the interfaces.
   
   We have synthesized superlattices with periods consisting of $n$ unit cells of LNO and 2 unit cells of SrMnO$_3$ (SMO).  SMO is an antiferromagnetic band insulator with Mn in a $3d^3$ ($t_{2g}^{3}$) electronic configuration.  Both SMO ($c = 3.805$ \AA) and LNO ($c = 3.83$ \AA) are under tensile strain when epitaxially deposited on SrTiO$_3$ ($c = 3.905$ \AA).  A combination of electron and x-ray scattering measurements confirm the (LNO)$_n$/(SMO)$_2$ superlattices are strained to the SrTiO$_3$ and exhibit high crystalline quality with abrupt interfaces.  As $n$ is increased from 1 to 2, the superlattices transition from gapped insulator to hopping conductor.  Upon increasing $n$ to 4, a metallic state is recovered although with a mean free path ten times less than that of pure LNO.

\section{Experimental}
   The superlattices were grown on insulating SrTiO$_3$ (STO) (0 0 1) substrates in a custom-designed molecular beam epitaxy system at the Center for Nanoscale Materials at Argonne National Laboratory.  The system is described in detail elsewhere.\cite{Santos}  Prior to the growth, the substrates were rinsed under deionizated water for 15 min, then etched in a commercial buffered oxide etchant for 20 s in order to form a TiO$_2$ terminated surface.  Following the etch process, trichloroethylene was used to remove organic contaminants from the substrate surface.  The prepared substrate was then loaded into the growth chamber and exposed to 2 x 10$^{-6}$ Torr of pure ozone for 3 - 5 hr at room temperature before deposition.  
   
   The superlattices were grown under the conditions that were found to produce the best quality, pure LNO and SMO films. Deposition was carried out at 550$^\circ$C in flowing ozone, with the growth chamber pressure fixed at 2 x 10$^{-6}$ Torr.  The samples were grown by depositing a single elemental layer at a time.\cite{Eckstein90,Locquet94,Vasko96}  Brief anneal periods (15 - 25 s) followed the completion of each metal-oxide layer.  For example, deposition of a SMO/LNO bilayer on SrTiO$_3$ would be shuttered as follows: TiO$_2$ (substrate) /SrO/anneal/MnO$_2$/anneal/LaO/anneal/NiO$_2$/anneal.  This block-by-block approach was found to produce a smoother LNO film than one grown by codeposition (opening both La and Ni shutters simultaneously) at 650$^\circ$C.  In all superlattices, the SMO layers were deposited first on the STO substrates.  The number of periods was chosen such that the total sample thickness was 200 - 220 \AA.
   
  A 173-\AA~thick film of LNO grown under these conditions exhibited a $c$-axis parameter of 3.807 $\pm$ 0.003 \AA, a rocking curve of 0.040$^\circ$, and resistivity values of 5.4 x 10$^{-5}$ and 2.5 x 10$^{-4}$ $\Omega$-cm at 5 and 300 K, respectively.  A 145-\AA~thick film of SMO exhibited a $c$-axis parameter of 3.777 $\pm$ 0.003 \AA, a rocking curve of 0.057$^\circ$, and resistivity values of 2.4 x 10$^{4}$ and 0.4 $\Omega$-cm at 80 and 300 K, respectively.  X-ray scans of the (1 0 1) diffraction peak confirm both films have in-plane lattice constants of $\sim$3.905 \AA~and thus are believed to be strained.  The measured properties of these films are comparable to those reported for LNO and SMO films and bulk crystals.\cite{Nikolaev,Chen99,Dobin03,Horiba07,Chmaissem,Chiorescu}
  
  X-ray scattering measurements were performed on a Phillips X'Pert diffractometer and a Bruker D8 diffractometer.  The wavelengths used for x-ray measurements were 1.5406 and 1.5418 \AA~ for diffraction and reflectivity, respectively.  X-ray reflectivity data was fit using commercial software (PANalytical X'Pert Reflectivity, version 1.1) that utilizes Parratt's dynamical formalism.  Quantum Design PPMS (with external electronics) and MPMS systems were used for in-plane resistivity and magnetometry measurements, respectively.  Resistivity was measured in the four-point probe geometry using indium dots to contact the films.
  
\section{Structural Properties}   
\begin{figure}
\includegraphics[width=2.4 in]{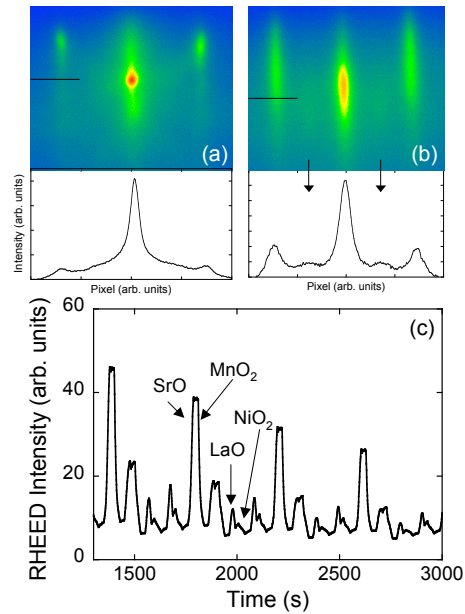}
\caption{(Color online) RHEED patterns before (a) and after (b) growth of a [(LNO)$_2$/(SMO)$_2$]$_{14}$ superlattice along with corresponding line profiles.  The short black lines indicate where each line profile was obtained.  The arrows in (b) highlight the spectral intensity believed to arise from a surface reconstruction of the NiO$_2$-terminated sample.  The specular spot intensity measured during deposition of the fourth through seventh periods of the superlattice is given in (c).}
\label{fig:rheed}
\end{figure}

  The superlattice growth process was monitored \textit{in situ} using reflection high energy electron diffraction (RHEED).  Figure~\ref{fig:rheed}(a,b) shows the RHEED pattern of the STO substrate before growth and the [(LNO)$_2$/(SMO)$_2$]$_{14}$ superlattice at the end of growth, respectively.  The streaky pattern shown in Fig.~\ref{fig:rheed}(b), consistent with a smooth sample surface, is representative of the superlattices deposited in this study.  The specular spot FHWM measured at the end of growth is 20$\%$ higher than that measured on the substrate.  The specular spot intensity oscillates in a periodic manner throughout the growth, as demonstrated in Fig.~\ref{fig:rheed}(c).  The RHEED intensity is maximized at the completion of SrO layers.  During deposition of the MnO$_2$ and NiO$_2$ layers, the RHEED intensity steadily decreases.  The oscillation amplitudes of the LaO and NiO$_2$ layers are less than those measured of the SrO and MnO$_2$ layers. This may arise from a mixed layer-by-layer and step-flow growth mode of the LaNiO$_3$ layers, as step-flow growth is known to produce negligible RHEED oscillations.\cite{Neave}   The RHEED intensity decreases with each period.  This decrease is a result of an accumulation of surface roughness with each superlattice cycle and a gradual decrease in emission current of the RHEED gun, which was not adjusted during deposition.
     
\begin{figure}
\includegraphics[width=2.4 in]{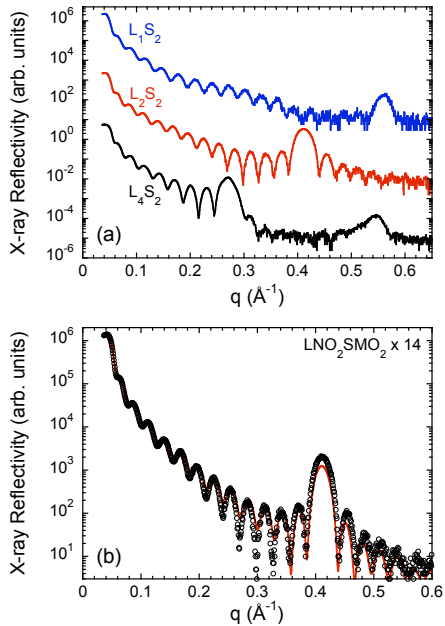}
\caption{(Color online) X-ray reflectivity of (LNO)$_n$/(SMO)$_2$ superlattices (a).  The notation above the Bragg reflections refers to the number of LNO (L) and SMO (S) unit cells in each superlattice period.  The fit obtained for the $n$ = 2 superlattice is shown in (b).}
\label{fig:xrr}
\end{figure}

   X-ray reflectivity, shown in Fig.~\ref{fig:xrr}(a), was used to determine the superlattice composition, thickness and interfacial roughness.  Strong Bragg reflections, arising from the difference in densities of LNO and SMO, are obtained indicating abrupt interfaces with intermixing limited to length scales of less than one unit cell.  
      
   A representative fit of the reflectivity is given in Fig.~\ref{fig:xrr}(b).  The fitting parameters are interfacial roughness, density of the LNO and SMO layers, and thickness of the LNO and SMO layers.  Interfacial roughness corresponds to the length scale over which the density changes from LNO to SMO.  Both interlayer mixing and layer morphology contribute to interfacial roughness.  For all superlattices, the reflectivity was best fit using models where the interfacial roughness increased with increasing distance from the substrate, consistent with the decreasing RHEED intensity previously discussed.  Typical roughness values for the first 10 - 14 bilayers are 1.5 - 3 \AA.  The four bilayers closest to the surface have maximum roughness values between 3 - 4.3 \AA.  The modeled reflectivity is most sensitive to these roughness values near the surface.  The layer densities were between 5.5 and 5.65 g/cm$^3$ for SMO and 7.2 and 7.25 g/cm$^3$ for LNO, within 3 $\%$ of the calculated values.
   
\begin{figure}
\includegraphics[width=3.2 in]{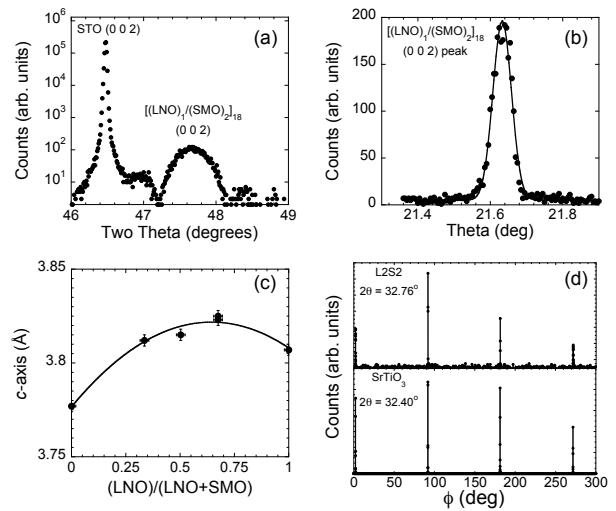}
\caption{2$\theta$-$\theta$ scan around the (0 0 2) peak of the [(LNO)$_1$/(SMO)$_2$]$_{18}$ superlattice (a).  A $\theta$ scan measured in the same superlattice at the (0 0 2) peak center is shown in (b).  The solid line is the Gaussian fit.  The average $c$-axis parameter of the superlattices is given in (c) as a function of the number of LNO unit cells divided by the total number of unit cells in each sample.  A $\phi$ scan measured at the (1 0 1) peak of a [(LNO)$_2$/(SMO)$_2$]$_{14}$ superlattice is shown in the top panel of (d), while the corresponding scan of the STO is shown in the bottom panel.}
\label{fig:xrd}
\end{figure}

   X-ray diffraction was used to investigate the lattice parameters and crystallinity of the superlattices.  Figure~\ref{fig:xrd}(a) shows a characteristic 2$\theta$-$\theta$ scan performed on a (0 0 2) peak.  The (0 0 2) peak was fit to a Gaussian function to obtain the peak center and full-width half maximum (FWHM).  The FWHM values, in units of momentum transfer, ranged from 0.026 - 0.039 \AA$^{-1}$ indicating that the peak widths are limited by the thickness of the superlattices.\cite{Warren}  The out-of-plane crystallinity (rocking curve) was measured by scanning $\theta$ at the (0 0 2) peak center, shown in Fig.~\ref{fig:xrd}(b).  The rocking curve widths ranged from 0.036 - 0.06$^\circ$, confirming the excellent out-of-plane crystallinity present in the superlattices.  For comparison, the rocking curves measured on the STO substrates ranged from 0.033 to 0.052$^\circ$.
   
   \begin{table*}
\caption{\label{tab:table1}Structural properties of the LNO and SMO films and (LNO)$_n$/(SMO)$_2$ superlattices (L$N$S2).  $R_S$ is the surface roughness, while $R_{Avg}$ is the average roughness of all interfaces in the superlattice.  Both $R_S$ and $R_{Avg}$ are obtained from x-ray reflectivity.  The error in the $c$-axis value is $\pm$ 0.005 \AA.  $\Delta$ is the error in period thickness compared to the targeted value.}
\begin{ruledtabular}
\begin{tabular}{cccccccc}
Sample&Thickness (\AA)&Superlattice Period (\AA)&$R_S$ (\AA)&$R_{Avg}$ (\AA)&$c$-axis (\AA)&$\Delta$ (\%)&Rocking Curve ($^\circ$)\\
\hline
LNO & 173 & - & 5.1 & - & 3.807 & - & 0.040\\
SMO & 145 & - & 4.5 & - & 3.777 & - & 0.057\\
L1S2 & 203 & 11.3 & 3.0 & 2.3 & 3.812 & -1.2 & 0.060\\
L2S2 & 215 & 15.4 & 3.7 & 2.3 & 3.815 & 0.8 & 0.050\\
L4S2 & 211 & 23.4 & 4.4 & 2.4 & 3.825 & 2.0 & 0.036\\
\end{tabular}
\end{ruledtabular}
\end{table*}

   The average $c$-axis lattice parameters obtained from the (0 0 2) peaks exhibit a non-linear dependence on the superlattice composition.  In all superlattices, the average $c$-axis parameter is larger than that of either the pure LNO or SMO films, as can be seen in Fig.~\ref{fig:xrd}(c).  The $c$-axis can be described by $c$(\AA) $= -0.011x^{2} + 0.14x + 3.777$, where $x$ is the volume of LNO in the superlattice divided by the total superlattice volume.   The increased $c$-axis parameter measured in the superlattices may result from a mixed valence state of interfacial Mn and Ni ions, leading to changes in their ionic radii. X-ray spectroscopy is anticipated to investigate the valence states of the transition metal ions.  

   The $c$-axis parameter can be compared with the superlattice period obtained from reflectivity in order to gauge the compositional error in the samples.  For example, the bilayer thickness of the [(LNO)$_2$/(SMO)$_2$]$_{14}$ superlattice is 15.38 \AA.  The thickness obtained by multiplying the average $c$-axis parameter by 4 is 15.26 \AA, yielding a difference between the measured and expected superlattice period of 0.8 $\%$.  In all (LNO)$_n$/(SMO)$_2$ superlattices, the difference between the measured and expected period ranges from -1.2 to 2.0 $\%$.
   
   The in-plane lattice constants and epitaxial alignment were investigated by performing $\phi$ scans on the (1 0 1) sample peak.  The (1 0 1) peaks were centered at 2$\theta$ values ranging from 32.7 to 32.8$^\circ$, indicating the superlattices are under tensile strain with in-plane lattice constants between 3.90 and 3.91 \AA.  $\phi$ scans yield four peaks separated by 90$^\circ$, as shown in Fig.~\ref{fig:xrd}(d).  The (1 0 1) peaks are commensurate with those from the STO substrate confirming the [1 0 0]$_{SL}$ $\parallel$ [1 0 0]$_{STO}$ epitaxial relation.  The widths of the $\phi$ scan superlattices peaks are approximately 0.1$^\circ$, compared to values of $\sim$0.04$^\circ$ measured on the STO.  
   
   Table I lists the structural properties of all samples used in this study.  In summary, the superlattices are highly crystalline, exhibit abrupt interfaces, and are strained with $c$/$a$ ratios of 0.978 $\pm$ 0.002.

\section{Transport Properties}      

\begin{figure}
\includegraphics[width=2.7 in]{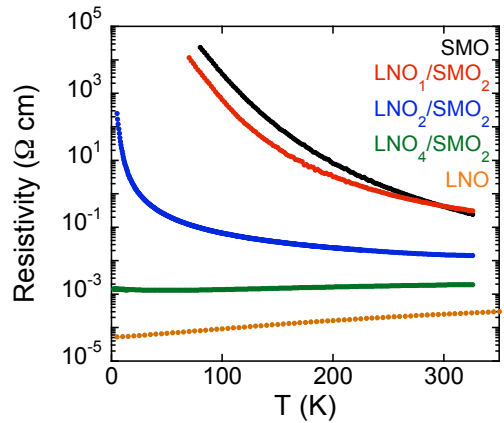}
\caption{(Color online) Resistivity in the superlattices and single material films.}
\label{fig:resistivity}
\end{figure}

   The temperature dependent resistivity ($\rho$) is given in Fig.~\ref{fig:resistivity} for the (LNO)$_n$/(SMO)$_2$ superlattices and pure LNO and SMO films.  A metal-insulator transition is observed with $\rho$ spanning over nine orders of magnitude between LNO and SMO. 
   
\begin{figure}
\includegraphics[width=2.1 in]{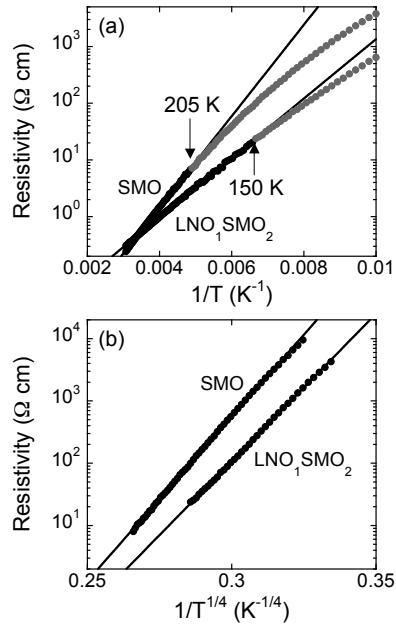}
\caption{High T activated behavior (a) and low T 3D variable range hopping (b) in SMO (90 - 200 K) and the $n$ = 1 superlattice (80 - 150 K).}
\label{fig:smo}
\end{figure}

   To gain an understanding of how the conduction mechanism evolves as a function of $n$, $\rho$ of the insulating samples was fit to a variety of models used to describe activated behavior, variable range hopping (VRH), and polaronic transport.  The resistivity of the SMO film and $n$ = 1 superlattice was best fitted to activated behavior at high T and 3D VRH at low T.  Figure~\ref{fig:smo}(a) shows the high T resistivity of the SMO film and (LNO)$_1$/(SMO)$_2$ superlattice fit to the activated form typical of gapped insulators, 
   \begin{equation}
  \rho=\rho_0\exp(E_A/k_BT).
  \label{eq:one}
\end{equation}
  An activation energy, $E_A$, of 160 meV is obtained from the SMO film between 205 and 325 K.  While this is larger than the value of 25 meV measured in bulk polycrystalline SMO,\cite{Chiorescu} it is in agreement with the band gap of 300 meV calculated for cubic SMO assuming $E_G$ $\sim$ 2$E_A$.\cite{Sondena}  The activation energy is reduced to 105 meV in the (LNO)$_1$/(SMO)$_2$ superlattice, indicating that the superlattice remains a gapped insulator, although with a narrower band gap than pure SMO.  As seen in Fig.~\ref{fig:smo}(b), at low T both samples are well described by VRH given by,
   \begin{equation}
    \rho=\rho_0\exp(T_0/T)^{\alpha},
    \label{eq:two}
    \end{equation}
    where $\alpha$ = 1/4 and 1/3 for 3D and 2D hopping, respectively.  The SMO and $n$ = 1 superlattice are best fit to the 3D behavior.  $T_0$ values of 2.1 x 10$^8$ and 1.4 x 10$^8$ K were obtained from the SMO and $n$ = 1 superlattice, respectively.  This transition from activated band conduction to hopping conduction between localized states within the gap is a common phenomenon in semiconductor transport.\cite{Shklovskii}
      
\begin{figure}
\includegraphics[width=3.3 in]{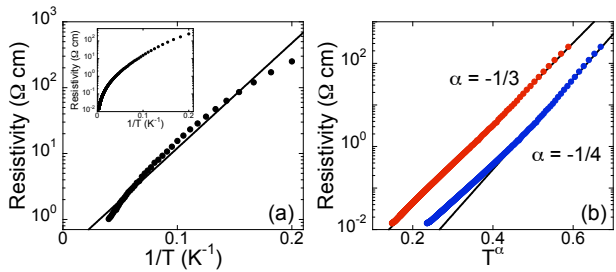}
\caption{(Color online) Resistivity of the [(LNO)$_2$/(SMO)$_2$]$_{14}$ superlattice fit to activated behavior from 5 - 25 K (a).  The inset of (a) shows $\rho$ as a function of 1/T over the range of 5 - 325 K.  The fits to 2D and 3D VRH are shown in (b) from 5 - 325 K.}
\label{fig:s2l2}
\end{figure}

  When a second LNO layer is added to the superlattice period, $\rho$ is greatly reduced compared to the $n$ = 1 superlattice.  Another difference between the two samples is that the $n$ = 2 superlattice does not exhibit activated behavior over any appreciable range of $\rho$ and T.  As shown in Fig.~\ref{fig:s2l2}(a), an activated model does not reproduce the data over any temperature range.  While both the 2D and 3D VRH models fit the low T data (5 - 25 K) equally well, the 2D VRH equation fits the data over the full 5 - 325 K range as can be seen in Fig.~\ref{fig:s2l2}(b).  The $T_0$ values for the 2D and 3D VRH fits are 2.6 x 10$^5$ and 3.9 x 10$^5$ K, respectively.  For both the 2D and 3D models, the hopping activation energy is much larger than $k_BT$ over the range of measurement temperatures, validating the use of VRH models.\cite{Shklovskii, Hammer}
  
  This result indicates that as the superlattice structure changes from $n$ = 1 to $n$ = 2 the band gap vanishes, or is at least substantially reduced.  It should be noted that in the $n$ = 1 superlattice each NiO$_2$ layer is between a LaO and a SrO layer, which may lead to an electronic configuration other than 3$d^7$ on the Ni.  In the $n$ = 2 superlattice, the layering is LaO/NiO$_2$/LaO/NiO$_2$/SrO and thus one of the NiO$_2$ planes has LaO planes on either side of it.  We speculate that a complete LaO/NiO$_2$/LaO structure is weakly conducting, although non-metallic.  In this view, the $n$ = 2 superlattice behaves like the sum of a strongly insulating and weakly insulating material, unlike the $n$ = 1 superlattice, which behaves like a gapped insulator. 

\begin{figure}
\includegraphics[width=2.2 in]{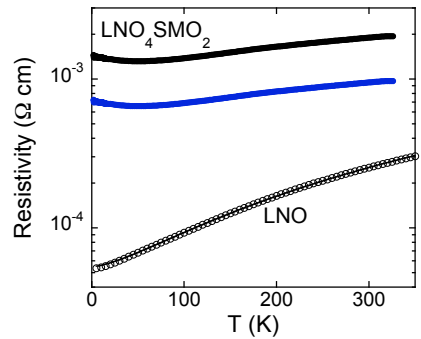}
\caption{(Color online) Resistivity of the [(LNO)$_4$/(SMO)$_2$]$_9$ superlattice and LNO film.  Shown in blue is the resistivity of the superlattice normalized by the number of LaO/NiO$_2$/LaO layers in the superlattice period (3/6).  The solid line shows the $T^{1.5}$ fit to the LNO resistivity.}
\label{fig:lno}
\end{figure}

  The LNO and $n$ = 4 samples are metallic. However, the magnitude and T dependence of $\rho$ is quite different in the two samples as can be seen in Fig.~\ref{fig:lno}.  The LNO resistivity is well described by $\rho=\rho_0+AT^{1.5}$, where $\rho_0$ = 54.8 $\mu\Omega$-cm and $A$ = 3.84 x 10$^{-2}$ $\mu\Omega$-cm/K$^{1.5}$.  A $T^{1.5}$ dependence has been observed in the resistivity of bulk and thin film LNO\cite{Xu93}, which is attributed to the presence of localized spin fluctuations,\cite{Inaba} such as those arising from Ni$^{2+}$ ions.\cite{Gayathri98}  A $\rho \sim T^2$ behavior has also been observed at low T in some samples,\cite{Xu93,Sreedhar} and is believed to be induced by oxygen vacancies.\cite{Gayathri98}  Our LNO film could not be fit to a $T^2$ dependence even from 5 - 50 K, suggesting a robust oxygen stoichiometry.  
    
  The resistivity of the $n$ = 4 superlattice increases from 1.32 to 1.9 m$\Omega$-cm as T is increased from 50 to 300 K.  In this temperature range, $\rho$ cannot be described by a single $\rho=\rho_0+AT^{m}$ equation.  Instead, two different behaviors are observed with $m$ = 1.9 from 70 to 170 K and $m$ = 1 from 180 to 300 K.  The mean free path ($l$) is estimated from Eq. (\ref{eq:three}):\cite{Lee85} 
\begin{equation}
  l = \frac{\hbar}{e^2} \frac{1}{\rho_0} \left(\frac{3\pi^2}{n^2}\right)^{1/3},
  \label{eq:three}
\end{equation}
 where $\rho_0$ is the residual resistivity and $n$ is the electron concentration, assumed to be 1.7 x 10$^{22}$ cm$^{-3}$.\cite{Gayathri98}  Using $\rho_0$ normalized by the volume of LaO/NiO$_2$/LaO layers in the superlattice (blue curve in Fig.~\ref{fig:lno}), $l$ is found to be 3 \AA, on the order of a single unit cell.  This yields a $k_Fl$ value of 2.4.  For comparison, $l$ of the LNO film is 35 \AA.  Finally, by treating the sample as nine equivalent resistors in parallel, the sheet resistance of each (LNO)$_4$ layer is 5.63 k$\Omega$/$\square$, or $0.22h/e^2$. 
 
  Below 50 K, the $n$ = 4 resistivity increases slightly ($\sim$ 10\%).  The origin of this increase is unknown as the low T resistivity cannot be fit to activated or hopping models.  We have attempted to fit the data to equations that describe electron-electron interactions in 2D and 3D.\cite{Lee85}  However, the results are inconclusive due to the limited temperature and resistivity range.  A negative magnetoresistance (MR) is observed below 100 K, suggesting spin-related scattering may contribute to the increased $\rho$.  The magnitude of the negative MR increases with decreasing T, reaching a maximum of 1.5 $\%$ at 5 K and 8 T.  The MR exhibits the same behavior when the field is applied in-plane and out-of-plane, with the field perpendicular to the current in both cases.  The isotropic MR is evidence that weak localization is not responsible for the increase in $\rho$ at low T.  A similar, negative MR was not observed in the LNO film.  

  Finally, we note that magnetization was measured as a function of field and temperature in all superlattices presented in this study.  No signs of ferromagnetism were observed within the noise of the measurements ($\sim$ 0.06 $\mu_B$/unit cell).  Neutron diffraction measurements are anticipated to determine if antiferromagnetic ordering is present in the superlattices.

\section{Conclusions}   
   Motivated by predictions of magnetic ordering and superconductivity, we synthesized epitaxial (LaNiO$_3$)$_n$/(SrMnO$_3$)$_2$ superlattices on STO substrates.  Using x-ray scattering, we confirmed the superlattices exhibit abrupt interfaces and are strained with $c$/$a$ $\sim$ 0.98.  While superconductivity and ferromagnetism are absent, the samples undergo a metal-insulator transition as $n$ is reduced from 4 to 2.  Both $n$ = 1 and 2 samples are insulating, however, they exhibit different transport behavior.  The $n$ = 1 sample acts as a gapped insulator, while the addition of a second LNO layer to each superlattice period ($n$ = 2) leads to hopping transport through non-gapped conduction channels.  
   
   A comparison to (LaMnO$_3$)/(SrMnO$_3$) superlattices is informative as both LaMnO$_3$ (LMO) ($t_{2g}^{3}e_{g}^{1}$) and LNO ($t_{2g}^{6}e_{g}^{1}$) have a lone $e_g$ electron.  In short period (LMO)/(SMO) superlattices, the $e_g$ electrons leak into the SMO layers giving rise to metallic behavior and robust magnetic ordering,\cite{Anand08, Salvador99, Yamada06, Adamo08} either ferro- or antiferromagnetic depending on the superlattice composition.  In contrast, the (LNO)$_1$/(SMO)$_2$ superlattice is strongly insulating suggesting that the SMO layers are significantly less doped than those in (LMO)$_1$/(SMO)$_2$ superlattices.\cite{Anand07}  Thus it appears that charge transfer is reduced in the nickelate/manganite superlattices compared to their all-manganite counterparts.  The structural integrity of the (LNO)/(SMO) superlattices allows for future neutron diffraction measurements  to look for antiferromagnetic order and resonant x-ray scattering techniques to investigate orbital and electronic interfacial effects.  
   
\section{Acknowledgments}   
   We thank Chris Leighton and Sam Bader for useful comments.  Work supported by UChicago Argonne, LLC, Operator of Argonne National Laboratory (ÒArgonneÓ). Argonne, a U.S. Department of Energy (DOE), Office of Science laboratory, is operated under Contract No. DE-AC02-06CH11357. We acknowledge use of the oxide-MBE, x-ray diffraction, transport and magnetic characterization facilities at the Center for Nanoscale Materials. Use of the Center for Nanoscale Materials was supported by the U. S. Department of Energy, Office of Science, Office of Basic Energy Sciences, under Contract No. DE-AC02-06CH11357.  S.M. and A.B. acknowledge support from the Digital Synthesis FWP, funded by DOE-BES.  


\end{document}